\documentclass[journal]{IEEEtran}

\usepackage{cite}

\ifCLASSINFOpdf
   \usepackage[pdftex]{graphicx}
   \graphicspath{{../pdf/}{../jpeg/}}
   \DeclareGraphicsExtensions{.pdf,.jpeg,.png}
\else
\fi
\UseRawInputEncoding
\usepackage{amsmath}
\usepackage{array}
\usepackage{standalone}
\usepackage{etoolbox}
\newtoggle{arXiv}
\newtoggle{bibrun}
\usepackage{newtxtext, newtxmath}
\usepackage{graphicx}
\usepackage{color, calc}
\usepackage{array, booktabs}
\newcolumntype{L}{>{$}l<{$}}
\newcolumntype{C}{>{$}c<{$}}
\newcolumntype{R}{>{$}r<{$}}
\usepackage{tikz, tikz-3dplot}
\usetikzlibrary{arrows, arrows.meta, calc, positioning}
\usetikzlibrary{decorations.pathmorphing}	
\tikzset{>=Stealth}
\usepackage{siunitx, physics}
\usepackage{enumitem}
\setlist[description]{labelindent=0pt, leftmargin=\parindent, font=\normalfont\itshape}
\usepackage{url}
\usepackage{subfigure}
\usepackage{textcomp}
\usepackage{eurosym}
\usepackage{xfrac}
\usepackage{pgfplots}
\usepackage{stmaryrd}
\pgfplotsset{compat=1.17}
\usepackage{makecell}

\begin{document}
%
\title{Portable MRI for major sporting events - a case study on the MotoGP World Championship}

\author{\IEEEauthorblockN{
		Jos\'e\,M.\,Algar\'{\i}n\IEEEauthorrefmark{1}$^,$\IEEEauthorrefmark{7},
		Teresa\,Guallart-Naval\IEEEauthorrefmark{2}$^,$\IEEEauthorrefmark{7},
		Enrique\,Gastaldi-Orqu\'{\i}n\IEEEauthorrefmark{3},
		Rub\'en\,Bosch\IEEEauthorrefmark{2},
		Francisco\,J.\,Lloris\IEEEauthorrefmark{4},
		Eduardo\,Pall\'as\IEEEauthorrefmark{3},
		Juan\,P.\,Rigla\IEEEauthorrefmark{2},
		Pablo\,Mart\'inez\IEEEauthorrefmark{2},
		Jos\'e\,Borreguero\IEEEauthorrefmark{2},
		Roberto\,Alamar\IEEEauthorrefmark{5},
		Luis\,Mart\'i-Bonmat\'i\IEEEauthorrefmark{6},
		Jos\'e\,M.\,Benlloch\IEEEauthorrefmark{1}, 
		Fernando\,Galve\IEEEauthorrefmark{1}, and
		Joseba\,Alonso\IEEEauthorrefmark{1}}
	
	\IEEEauthorblockA{\IEEEauthorrefmark{1}MRILab, Institute for Molecular Imaging and Instrumentation (i3M), Spanish National Research Council (CSIC) and Universitat Polit\`ecnica de Val\`encia (UPV), 46022 Valencia, Spain}\\
	\IEEEauthorblockA{\IEEEauthorrefmark{2}Tesoro Imaging S.L., 46022 Valencia, Spain}\\
	\IEEEauthorblockA{\IEEEauthorrefmark{3}La Salud Hospital, Department of Orthopedic Surgery, 46021 Valencia, Spain}\\
	\IEEEauthorblockA{\IEEEauthorrefmark{4}PhysioMRI Tech S.L., 46022 Valencia, Spain}\\
	\IEEEauthorblockA{\IEEEauthorrefmark{5}Ricardo Tormo Circuit, 46380 Cheste, Spain}\\
	\IEEEauthorblockA{\IEEEauthorrefmark{6}Hospital Universitari i Polit\`ecnic La Fe, Medical Imaging Department, 46026 Valencia, Spain}\\
	\IEEEauthorblockA{\IEEEauthorrefmark{7}These authors contributed equally}\\
	
\thanks{Corresponding author: J. Alonso (joseba.alonso@i3m.upv.es).}}


\maketitle

\begin{abstract}
\newline
Purpose: {\normalfont The goal of this work is to showcase the clinical value that portable MRI can provide in crowded events and major sports competitions.}
\\
Methods: {\normalfont We temporarily installed a low-field and low-cost portable MRI system for extremity imaging in the medical facilities of the Ricardo Tormo Motor Racing Circuit during the four days of the Motorcycle Grand Prix held in Valencia (Spain), which closed the 2022 season of the MotoGP, Moto2 and Moto3 World Championships and the MotoE World Cup. During this time, we scanned 14 subjects, running a total of 21 protocols for wrist, knee and ankle imaging. Each protocol included a minimum of one T1-weighted 3-dimensional Rapid Acquisition with Refocused Echoes sequence for general anatomical information, and one 3D Short Tau Inversion Recovery sequence to highlight fluid accumulation and inflammation.}
\\
Results: {\normalfont The circuit medical staff were able to visualize a number of lesions and conditions in the low-field reconstructions, including gonarthrosis, effusion, or Haglund's syndrome, as well as metallic implants and tissue changes due to surgical interventions. Out of eight low-field acquisitions on previously diagnosed lesions, only two (a meniscus tear and a Baker cyst) were not detected by the experts that evaluated our images. The main highlight was that a low-field MRI scan on a subject reporting pain in a wrist revealed a traumatic arthritis which an X-ray radiograph and visual inspection had missed.}
\\
Conclusions: {\normalfont We have operated in a scenario where high-field MRI is unlikely to play a role but where a low-field system can lead to improved medical attention. Arguably, this can be extrapolated to numerous other environments and diverse circumstances. In the case reported here, system transport, installation in the circuit facilities and calibration were all uncomplicated. The images presented to the medical staff were mostly unprocessed and there is thus room for improvement. In conclusion, this work supports the claim that low-field MRI can likely provide added value whenever concepts such as accessibility, portability and low-cost outweigh exquisite detail in images.}
\end{abstract}


 \ifCLASSOPTIONpeerreview
 \begin{center} \bfseries EDICS Category: 3-BBND \end{center}
 \fi
%
\IEEEpeerreviewmaketitle


\section{Introduction}\label{sec:Intro}

\IEEEPARstart{T}{he} rather undemanding hardware and infrastructural requirements associated to low-field magnetic resonance imaging (LF-MRI) make it an ideal platform to expand the accessibility and applications of MRI beyond the restrictive environment of radiology departments in large clinical centers \cite{Webb2023,Cho2023,Marques2019,Sarracanie2020,Wald2020,Arnold2022}. LF-MRI scanners can be designed to be inexpensive and small because the main evolution field can be generated by permanent or resistive, rather than superconducting, magnets. The main price to pay for lowering the magnetic field strength $B_0$ is a loss in signal-to-noise ratio (SNR), but the LF-MRI community has devised strategies to mitigate the consequences and obtain clinically valuable reconstructions in multiple scenarios where standard high-field MRI is seldom (if ever) used. Recent accomplishments in this regard include in-hospital applications such as bedside and point-of care scans or in intensive care units \cite{Turpin2020,Sheth2021,Mazurek2021,Liu2021,Sheth2022,Yuen2022,Padormo2022}, as well as out-of-hospital scans with scanners mounted on vehicles \cite{Nakagomi2019,Deoni2022,Ikeda2022}.

The above achievements are all with permanent magnets where sizable ferromagnetic yokes guide the field lines through a magnetic circuit. Yokeless magnets furthermore allow for lightweight and portable designs \cite{Raich2004,Blumler2015,OReilly2020,Guallart2022}. The scope of applications enabled by truly portable MRI technologies is immense and largely unexplored. Novel applications include outpatient services in hospitals, use in emergency rooms, residential and hospice care, small clinics, rural areas, penitentiaries, sports clubs, school facilities, etc. Even outdoor uses become realistic \cite{Guallart2022}, e.g. in field hospitals, NGO and military camps, sports events, etc.

We recently demonstrated the capabilities of a new low-field extremity scanner, designed to be extremely portable and which was used indoors, outdoors and at a patient's residence \cite{Guallart2022,Guallart2022b}. In this paper, we study the potential MR value of our low-field extremity scanner for use in major sporting events, specifically in the Valencian Motorcycle Grand Prix held in the Ricardo Tormo Racing Circuit in Cheste (Spain) between November 3$^\text{rd}$ and 6$^\text{th}$, 2022. Overall, fourteen subjects were scanned in four days, including racers (1), track marshals (2), medical staff (3), safety car drivers (1), helicopter pilots (1), event organization staff (1), race control staff (1) and engineers (4). This work shows that a number of musculo-skeletal conditions can be diagnosed with MRI scanners operating at less than 0.1~T.

\section{Methods}\label{sec:Methods}
\subsection{Scanner \& acquisition}
\begin{figure*}
	\centering
	\includegraphics[width=1.95\columnwidth]{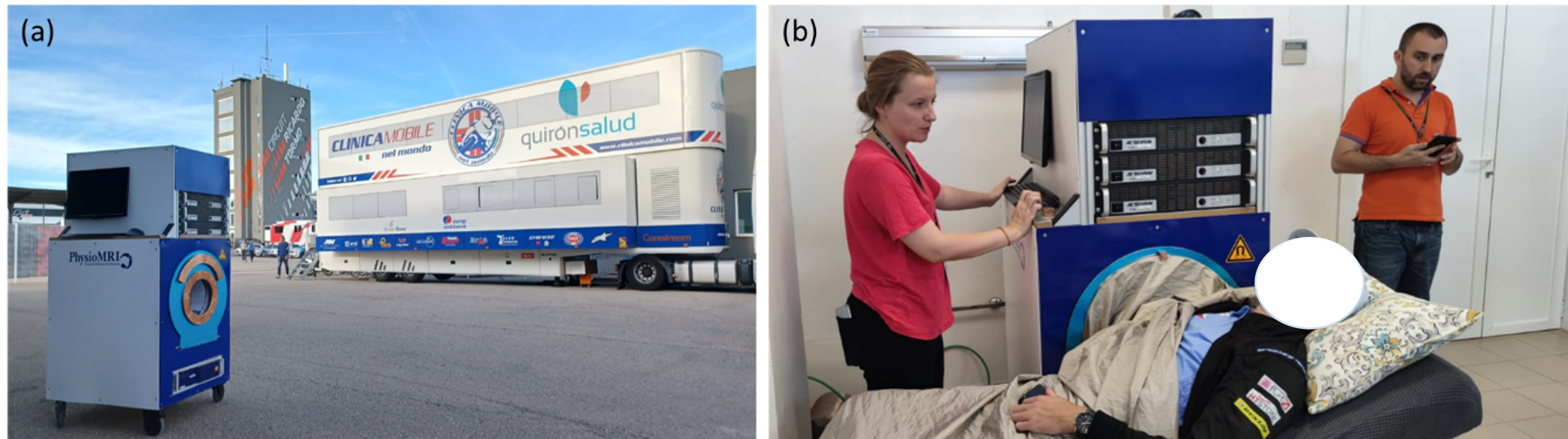}
	\caption{(a) Portable low-field extremity scanner in the paddock of the Ricardo Tormo Racing Circuit in Cheste (Spain). (b) Upper limb scan of a professional racer in the circuit medical facilities.}
	\label{fig:scanner}
\end{figure*}

Imaging was performed with a portable scanner (Fig.~\ref{fig:scanner}, \cite{Guallart2022}) based on a Halbach magnet made with a discrete array of around 5,000 NdFeO magnets, for a $B_0$ of around 72~mT homogeneous down to 3,000~ppm over a spherical field of view of 20~cm and 75~ppm for 10~cm. The complete system, including electronics and the wheeled mounting structure, weighs $<250$~kg and runs from a standard wall power outlet. We drive the scanner with MaRCoS, an inexpensive MAgnetic Resonance COntrol System able to accurately handle arbitrary waveforms \cite{Guallart2022b, Negnevitsky2022}. Further details on the scanner can be found in Refs.~\cite{Guallart2022,Guallart2022b}.

\begin{table*}
	\caption{Image acquisition parameters. All images have been acquired  with an echo train length of 5, echo spacing of 20 ms and effective echo time of 20 ms. 3D-STIR images in figure 2 (3) was obtained with inversion time of 90 (100) ms.  All images were reconstructed with partial Fourier methods, with a filling factor of 65\,\% along slice direction and setting to zero the points not acquired.}
	\centering
	\begin{tabular}{c c c c c c c c c}
		\hline
		Figure & \makecell{Anatomy \\ (pathology)} & Sequence & \thead{FOV \\ (mm$^3$)} & \# pixels & \thead{BW \\ (kHz)} & \thead{TR \\(ms)} & Avgs. & \thead{Scan time \\(min)} \\
		\hline
		
		\hline
		\ref{fig:wrists}(a) & \makecell{Wrist \\ (synovitis)} & 3D-RARE & $165\times120\times85$ & $120\times100\times14$ & 30 & 200 & 9 & 5.5 \\
		\hline
		
		\ref{fig:wrists}(b) & \makecell{Wrist \\ (synovitis)} & 3D-STIR & $165\times120\times85$ & $100\times80\times12$ & 25 & 800 & 9 & 15 \\
		\hline
		
		\ref{fig:wrists}(d) & \makecell{Wrist \\ (healthy)} & 3D-STIR & $165\times120\times85$ & $100\times80\times12$ & 25 & 800 & 9 & 15 \\
		\hline
		
		\hline
		\ref{fig:knees}(a) & \makecell{Knee \\ (implant)} & 3D-RARE & $150\times150\times180$ & $120\times120\times50$ & 30 & 200 & 8 & 20.8 \\
		\hline
		
		\ref{fig:knees}(b) & \makecell{Knee \\ (ligaments)} & 3D-RARE & $180\times150\times150$ & $120\times120\times28$ & 30 & 200 & 4 & 5.8 \\
		\hline
		
		\ref{fig:knees}(c) & \makecell{Knee \\ (joint effusion)} & 3D-STIR & $180\times150\times150$ & $100\times100\times22$ & 25 & 800 & 4 & 15.3 \\
		\hline
		
		\ref{fig:knees}(d) & \makecell{Knee \\ (meniscus fracture)} & 3D-RARE & $180\times150\times150$ & $120\times120\times28$ & 30 & 200 & 5 & 7.3 \\
		\hline
		
		\ref{fig:knees}(e) & \makecell{Knee \\ (baker cyst)} & 3D-STIR & $100\times150\times150$ & $100\times100\times28$ & 25 & 800 & 3 & 14.6 \\
		\hline
		
		\ref{fig:knees}(f) & \makecell{Knee \\ (joint effusion)} & 3D-RARE & $180\times150\times161$ & $120\times120\times28$ & 30 & 200 & 4 & 5.8 \\
		\hline
		
		\ref{fig:knees}(g) & \makecell{Knee \\ (healthy)} & 3D-RARE & $100\times150\times150$ & $120\times120\times28$ & 30 & 200 & 4 & 5.8 \\
		\hline
		
		\ref{fig:knees}(h) & \makecell{Knee \\ (healthy)} & 3D-STIR & $100\times150\times150$ & $100\times100\times28$ & 25 & 800 & 3 & 14.6 \\
		\hline
		
		\hline
		\ref{fig:ankles}(a) & \makecell{Ankle \\ (Haglund's)} & 3D-RARE & $180\times150\times132$ & $120\times90\times14$ & 30 & 200 & 10 & 5.5 \\
		\hline
		
		\ref{fig:ankles}(b) & \makecell{Ankle \\ (Haglund's)} & 3D-RARE & $180\times150\times104$ & $120\times90\times14$ & 30 & 200 & 10 & 5.5 \\
		\hline
		
		\ref{fig:ankles}(c) & \makecell{Ankle \\ (implant)} & 3D-RARE & $189\times150\times103$ & $80\times60\times6$ & 20 & 200 & 42 & 6.6 \\
		\hline
		
		\ref{fig:ankles}(d) & \makecell{Ankle \\ (healthy)} & 3D-RARE & $180\times150\times110$ & $120\times90\times14$ & 30 & 200 & 10 & 5.5 \\
		\hline
		
	\end{tabular}
	\label{tab:seq_params}
\end{table*}

The pulse sequences and parameters employed for the acquisitions during the MotoGP event are provided in Table~\ref{tab:seq_params}. We used two families of sequences: T1-weighted 3-dimensional Rapid Acquisition with Refocused Echoes (RARE, \cite{Hennig1986}), and 3D Short Tau Inversion Recovery fast spin-echo (fast STIR, \cite{Thorpe1994}). We acquired images of the subjects' wrists (1 injured, 4 healthy), knees (6 injured, 4 healthy) and ankles (3 injured, 3 healthy) for a total of 21 acquisitions on 14 subjects. All images are reconstructed with Fast Fourier Transform protocols and the only data processing is basic Block-Matched 4D filtering \cite{Maggioni2013}.

\subsection{System installation}
The system was transported in a small truck, installed in the main surgery room of the circuit medical facilities (Fig.~\ref{fig:scanner}b), and operational around 30 minutes after arrival. We encountered no setbacks for reaching normal working conditions; the RF circuitry simply required small adjustments to the impedance and matching electronics due to the slightly higher temperature of the room with respect to our laboratories.

We found the surgery room to be rather clean in the relevant bandwidth of the electromagnetic spectrum (around 3~MHz), and the grounded conductive cloth we typically use (Fig.~\ref{fig:scanner}b) sufficed to bring the MR signal noise down to Johnson-compatible levels. This was the case even on Sunday (race day), where stronger electromagnetic interference was visible without the Faraday shield, presumably due to the presence of the Spanish National Police, the Spanish Army, media, helicopters (the circuit medical facilities were around 50~m away from the heliport), and close to 100,000 spectators sitting in the bleachers.

The scanner is still in a prototype phase and has not been submitted for Medical Device Regulation (MDR) approval, so it was used only after explicit informed consent was provided by the scanned volunteers, and it did not interfere with the standard clinical protocols approved by the MotoGP organization and the circuit medical direction. The circuit medical staff were informed that the portable MRI scanner was available for their use, and it was operated by the trained staff of the MRILab when required by the medics.

\subsection{Radiological analysis}

All acquired images have been evaluated by expert radiologists and orthopedic surgeons. The images were provided as NeuroImaging Informatic Technology Initiative (NIFTI) files, along with details on the pulse sequences employed, the symptoms reported by the injured patients and, where available, previously existing images and diagnoses.

\section{Results}\label{sec:Results}
\subsection{Wrist scans}\label{sec:wrists}
\begin{figure*}
	\centering
	\includegraphics[width=1.8\columnwidth]{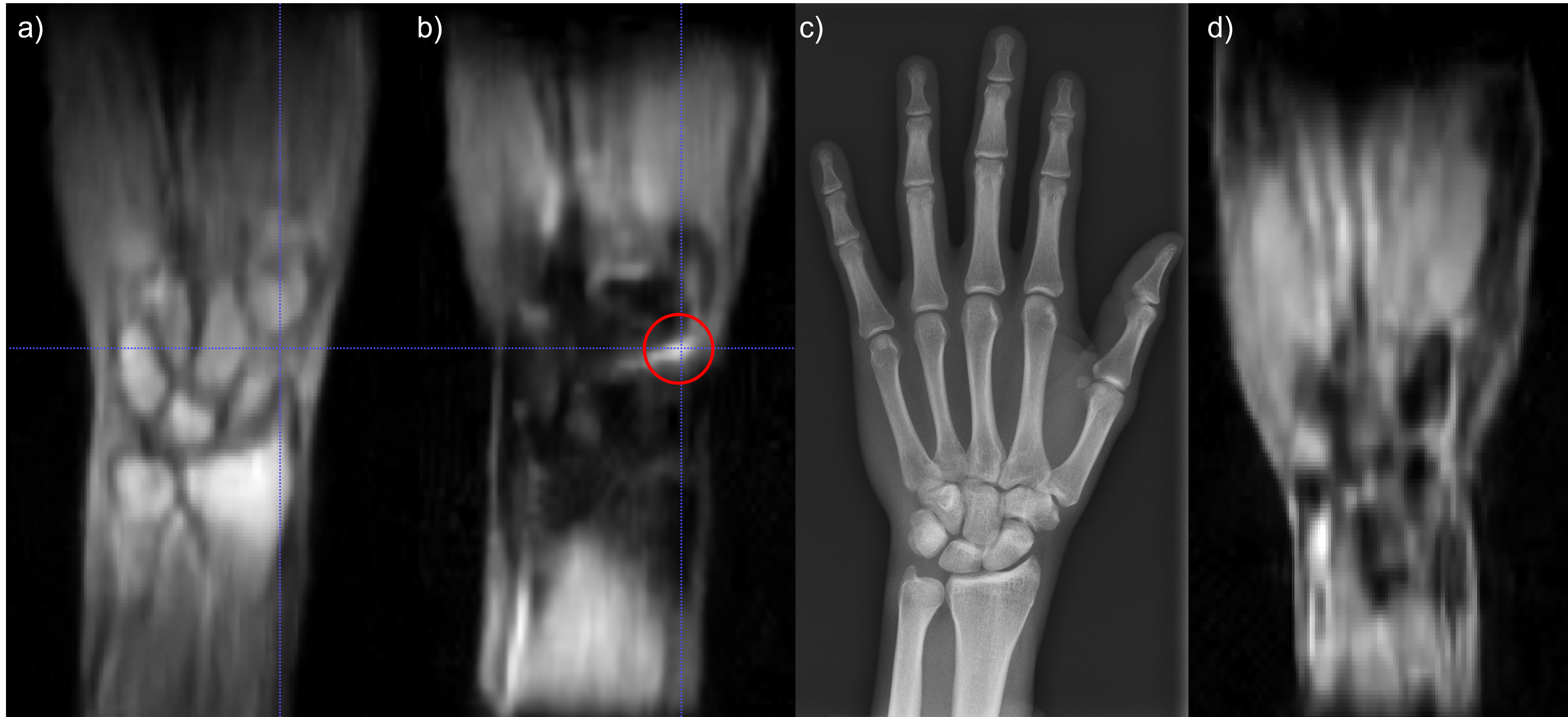}
	\caption{(a) 3D-RARE (T1-weighted) scan of a subject's right wrist. The image shows the main anatomic features but no lesion is visible. (b) 3D-STIR scan of the same wrist. The circled bright region indicates inflammation between the scaphoid, trapezium and trapezoid bones, possibly a traumatic arthritis. (c) The lesion is invisible in the X-ray image. (d) 3D-STIR acquisition on a healthy subject, revealing no lesion.}
	\label{fig:wrists}
\end{figure*}

Figure \ref{fig:wrists} shows images of the right wrist of a subject who had suffered an accident two weeks before the race and reported pain. Following the established protocols, the subject was scanned with an X-ray radiograph (Fig.~\ref{fig:wrists}c), which revealed no lesion. The subject was then scanned in our low-field system, where a 3D T1-weighted RARE acquisition also showed no anomaly (Fig.~\ref{fig:wrists}a). The subsequent STIR scan featured a bright volume between the scaphoid, trapezium and trapezoid bones (Fig.~\ref{fig:wrists}b), indicating a possible traumatic arthritis as judged by the orthopedist in charge at the medical center. For comparison, we scanned the right wrists of four healthy volunteers. None of them showed a bright region between the wrist bones (Fig.~\ref{fig:wrists}d).

\subsection{Knee scans}\label{sec:knees}
\begin{figure*}
	\centering
	\includegraphics[width=2\columnwidth]{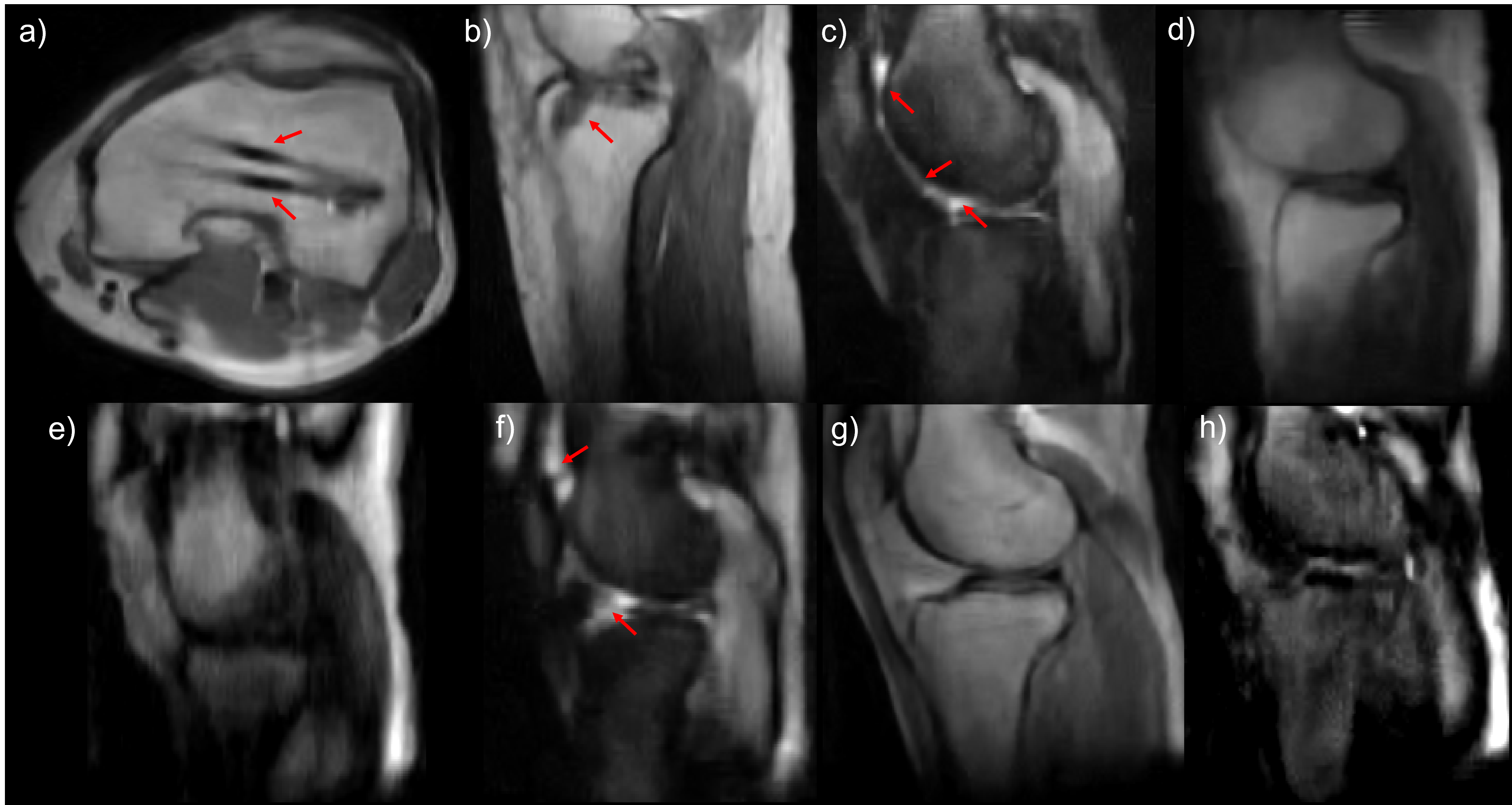}
	\caption{a) Femoral shaft osteotomy (RARE), b) intervention at the tibial tunnel of the anterior cruciate ligament and the femoral insertion of the posterior (RARE), c) fluid build-up due to joint effusion (STIR), d) meniscus fracture (not visible, RARE), e) Baker cyst (not visible, STIR), f) fluid build up (STIR), g) healthy (RARE), h) healthy (STIR).}
	\label{fig:knees}
\end{figure*}

Figure \ref{fig:knees} shows images of multiple knee acquisitions. We scanned six knees with different pathologies, some of them diagnosed before the MotoGP event. The circuit medical staff was able to identify a femoral shaft osteotomy in one knee (Fig.~\ref{fig:knees}a), the results of an intervention on anterior cruciate ligament and the femoral insertion of the posterior (Fig.~\ref{fig:knees}b) and fluid build-up due to joint effusion in two volunteers reporting pain (Fig.~\ref{fig:knees}c and f). The images in Fig.~\ref{fig:knees}d and e correspond to volunteers with a meniscus fracture and a Baker cyst, respectively. These were previously diagnosed on a 1.5 T scanner, but not visible in our low-field system with the employed protocol (note that the cyst may have decreased in size and even disappeared since it was diagnosed, and they are usually best visualized in T2-weighted images). In addition, we scanned four healthy knees, which showed no anomalies in the reconstructions.

\subsection{Ankle scans}\label{sec:ankles}
\begin{figure}
	\centering
	\includegraphics[width=1\columnwidth]{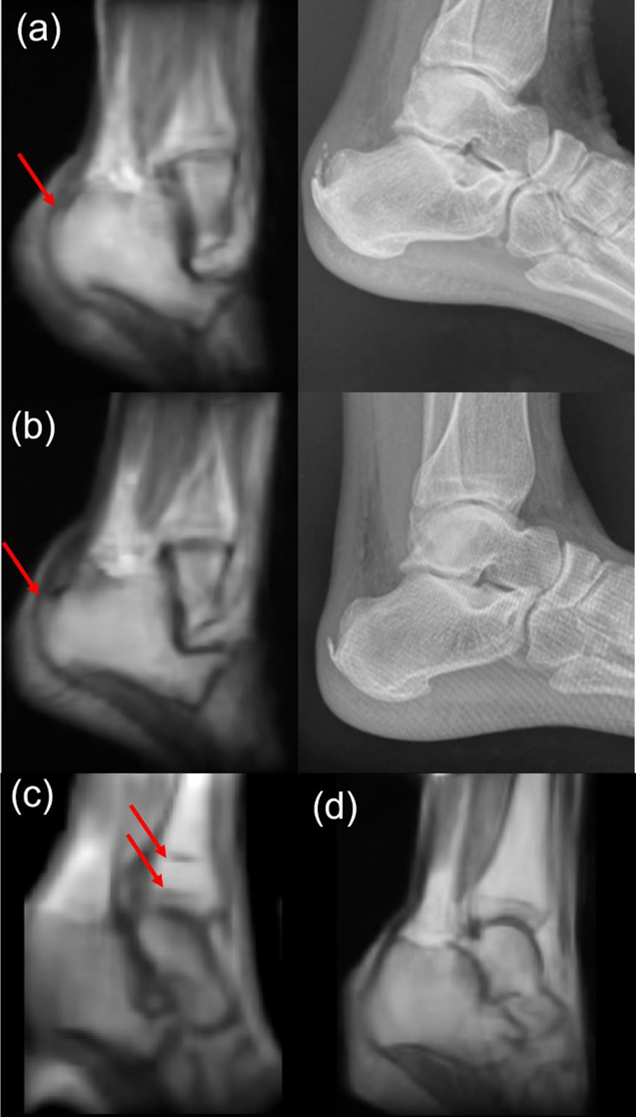}
	\caption{Low-field MR image (left) and X-ray radiograph (right) of a right (a) and left (b) ankle with Haglund's syndrome and intra-tendineal calcification of the Achilles tendon (red arrows). c) MRI scan of an operated ankle, with screws and plates indicated with red arrows. d) MRI scan of a healthy ankle.}
	\label{fig:ankles}
\end{figure}

Figure \ref{fig:ankles} shows scans of four ankles. The first volunteer suffered Hanglund's deformities (retrocalcaneal exostosis) and intra-tendineal calcification of the Achilles tendon. Both are evident in the X-ray radiograph (right), and visible in the low-field MRI scan in Fig.~\ref{fig:ankles}a and b. The ankle in Fig.~\ref{fig:ankles}c suffered a bone fracture and was fixed with metallic screws and plates, which appear free of artifacts in the low-field reconstructions. In addition to the scans shown in the figure, we performed two scans of healthy volunteers (Fig.~\ref{fig:ankles}d).

\section{Discussion}\label{sec:Disc}

The results of our work demonstrate the potential of portable MRI to improve medical protocols in major sporting events. While low-field MRI scanners have some drawbacks, such as reduced SNR and poorer resolution compared to conventional high-field MRI, we were still able to identify several conditions using a  protocol with only T1-weighted and inversion recovery images. Specifically, we were able to identify traumatic arthritis (Fig.~\ref{fig:wrists}b), the results of osteotomy (Fig.~\ref{fig:knees}a) and cruciate ligament interventions (Fig.~\ref{fig:knees}b), joint effusion (Fig.~\ref{fig:knees}c and f), Haglund's syndrome (Fig.~\ref{fig:ankles}a and b), calcification of the Achilles tendon (Fig.~\ref{fig:ankles}b), as well as metallic screws and plates (Fig.~\ref{fig:ankles}c).

Figure \ref{fig:knees}e shows images of a volunteer previously diagnosed with a Baker cyst using a 3 T MRI scanner. However, we were unable to detect the Baker cyst with our protocol, as this pathology is typically diagnosed using T2-weighted images, especially in the axial plane, where the fluid-distended cyst is easily depicted \cite{Perdikakis2013}. Including T2-weighted images in the protocol could potentially help detect this and other injuries.

We also acquired images of the knee of a volunteer previously diagnosed with a meniscus tear (Fig.~\ref{fig:knees}d). However, none of the doctors who reviewed the images were able to identify the injury. Meniscus tears are typically reflected in MRI images as small bright regions inside the menisci due to liquid insertion in the fracture (T1, IR, T2). The images obtained with our system did not show this brightness, likely due to the higher resolution required to capture small regions. This suggests that small lesions can be missed. Increasing the resolution at the expense of making the protocol longer may help to detect small injuries, although trade-offs between resolution and acquisition time would need to be carefully considered.

The ankle images in Fig.~\ref{fig:ankles} are not purely sagittal. The orientation employed provides a useful view of the subtalar joint, but hampers diagnosis of lesions in the calcaneous and the Achilles tendon, which was the consulted physicians disliked. This can be addressed by properly angulating the images, which we have started to implement in MaRCoS.

\section{Conclusion}\label{sec:Concl}

In this study, we present the first MRI scans taken directly at a major sporting event (2022 MotoGP, Moto2 and Moto3 World Championships, and the MotoE World Cup). Our findings demonstrate the potential of portable low-field MRI scanners in such scenarios, where traditional MRI is not part of the protocols due to the unavailability of MRI scanners in small medical facilities.

We utilized our portable and low-cost MRI system \cite{Guallart2022} to perform the scans. The setup process was straightforward and required only minor adjustments to the RF circuitry due to frequency drift caused by temperature changes during transport and installation. The scanner was ready for use merely 30 minutes after arriving at the circuit medical facilities, demonstrating the ease of operation of the system. The time needed to set up the system after arrival at the medical facilities can be further reduced by incorporating electronics that can automatically tune and match the RF coils \cite{Sohn2015}.

Our results demonstrate that LF-MRI scans can provide valuable information in the diagnosis and monitoring of injuries in sporting events. Some pathologies may still require an improved scanner performance to be detected. For example, we were able to detect traumatic arthritis in the wrist that would have otherwise gone unnoticed by the MotoGP medical staff, but we were not able to visualize meniscus tears, which need high-resolution images, or a Baker cyst, which may require T2-weighted images in the protocol. We expect the system to be useful for diagnosing other musculo-skeletal pathologies, including lesions in the Achilles tendon (aside from calcifications), in the plantar fascia ligament, anterior and posterior cruciate ligament injuries, patellofemoral misalignment, bone edemas, subtalar injuries, and astragalus fracture, among others.

Future work can aim to improve performance in several ways. Increasing the resolution with longer scan times is an option, even if it can lead to patient discomfort and increased probability of motion artifacts. An alternative approach to increase the resolution is through post-processing. For example, convolutional neural networks have demonstrated great potential in increasing the resolution of these scans \cite{DeLeeuwdenBouter2022}. We expect that the diagnostic potential of the system can be significantly boosted by machine learning algorithms and advanced data processing techniques which we have not used in this work. Sequences that optimize the SNR per unit time, such as MR fingerprinting or steady-state free precession, may also help \cite{Marques2019}.

Considering that our scanner has been specifically designed for extremity imaging, designing pulse sequences for hard tissue imaging, such as bone, tendons, and ligaments, is of particular interest. Zero Echo Time (ZTE) sequences, such as Pointwise-Encoding Time reduction with Radial Acquisition (PETRA \cite{Weiger2012, Grodzki2012}), have shown promising results in low-field settings \cite{Algarin2020}. Spin-locked pulse sequences have also been effective in selectively exciting slices while preserving signal from hard tissues \cite{Borreguero2023}, enabling 2D acquisitions on ZTE-like sequences for hard tissues.

Another area of improvement for our scanner is related to the conductive blanket. Covering the volunteers with the conductive blanket allowed us to prevent electromagnetic interference and reduce the noise picked up by the radiofrequency coil to the Johnson limit. However, the blanket can be uncomfortable for the subject and raise hygiene concerns. Active noise cancellation methods may offer a solution to these issues \cite{Srinivas2021, Liu2021, Parsa2023}.

\section{Contributions}

Low-field images taken by JMA and TGN with help from RB and JA. Portable system upgraded by TGN, JMA, RB, FJL, EP, JPR, PM, JB, FG and JA. Data analysis and evaluation by JMA, TGN and JA. Images assessed by EG and LMB. Portability experiments conceived by JA and RA. Project conceived and supervised by JMB, FG and JA. Paper written by JA, JMA and LMB, with input from all authors.

\section*{Acknowledgment}

We thank MD Vicente Vila and MD \'Angel Charte for their help organizing the installation of the portable MRI system for the Valencian MotoGP event. This work was supported by the Ministerio de Ciencia e Innovaci\'on of Spain through research grant PID2019-111436RB-C21, the European Union through the Programa Operativo del Fondo Europeo de Desarrollo Regional (FEDER) of the Comunitat Valenciana (IDIFEDER/2021/004), the Generalitat Valenciana (CIPROM/2021/003), and the Agencia Valenciana de la Innovaci\'on (INNVA1/2022/4).

\section*{Ethical statement}
All participants in this work were adults and provided written informed consent for this study. Ethical approval was obtained from the Ethics Committee (CEIm) of La Fe Hospital in Valencia (CEIm-F-PE-01-16, research agreement number 2022-187-1).

\section*{Conflict of interest}
PhysioMRI Tech S.L. is a for-profit organization spun off the Institute for Molecular Imaging and Instrumentation and proprietor of the low-field scanner presented in this work. JMA, JB, JMB, FG and JA have patents pending that are licensed to PhysioMRI Tech S.L. JMA, JMB, FG and JA are co-founders of PhysioMRI Tech S.L. FJL is an employee at PhysioMRI Tech S.L. All other authors declare no competing interests.

\ifCLASSOPTIONcaptionsoff
  \newpage
\fi


\end{document}